%% file: main.tex
\documentclass[acmsmall]{acmart}

\AtBeginDocument{%
  }

\usepackage{amsfonts}
\usepackage{graphicx}
\usepackage{array}
\usepackage{url}
\usepackage{fancybox}
\usepackage{multirow}
\usepackage{color}
\usepackage{colortbl}
\usepackage{balance}
\usepackage{fancyhdr}
\usepackage{subfig}
\usepackage{diagbox}
\usepackage{bbding}
\usepackage{placeins}
\usepackage{booktabs}
\usepackage{tabularx}
\usepackage{enumitem}
\usepackage{xcolor,lipsum}
\usepackage{textcomp}
\usepackage{algorithmic}
\usepackage{listings}
\usepackage{color}
\usepackage{commath}

\usepackage{bm}
\usepackage{amsmath,amsfonts}
\usepackage{algorithmic}
\usepackage{tcolorbox}
\usepackage{graphicx}
\usepackage{listings}
\usepackage{makecell}
\usepackage{adjustbox}
\usepackage{multirow}
\usepackage{graphicx}
\usepackage{svg}
\usepackage{xspace}
\usepackage{dsfont}
\usepackage{lipsum}
\usepackage{centernot}
\usepackage{diagbox}
\usepackage{tabularx}

\newcolumntype{P}[1]{>{\centering\arraybackslash}p{#1}}
\usepackage[ruled,vlined,linesnumbered]{algorithm2e}

\SetCommentSty{mycommfont}
\usepackage{pifont}
\usepackage{enumitem}
\usepackage{caption}
\usepackage{geometry}
\usepackage{pgf-pie}
\usepackage{multicol}
\usepackage{listings}
\usepackage{ragged2e}

\definecolor{light-gray}{rgb}{.906,  .902,  .902}

\newcommand{\toolname}{\textsc{Code2API}\xspace}

\setcopyright{rightsretained}
\acmDOI{10.1145/3660811}
\acmYear{2024}
\copyrightyear{2024}
\acmSubmissionID{fse24main-p946-p}
\acmJournal{PACMSE}
\acmVolume{1}
\acmNumber{FSE}
\acmArticle{104}
\acmMonth{7}
\received{2023-09-28}
\received[accepted]{2024-04-16}

\begin{document}
\title[Generating Reusable APIs with CoT and In-Context Learning]{Are Human Rules Necessary? Generating Reusable APIs with CoT Reasoning and In-Context Learning}


\author{Yubo Mai}
\affiliation{%
\institution{The State Key Laboratory of Blockchain and Data Security, Zhejiang University}
\country{China}
}
\email{12021077@zju.edu.cn}

\author{Zhipeng Gao}
\authornote{This is the corresponding author}
\affiliation{%
\institution{Shanghai Institute for Advanced Study of Zhejiang University}
\country{China}
}
\email{zhipeng.gao@zju.edu.cn}

\author{Xing Hu}
\affiliation{%
\institution{The State Key Laboratory of Blockchain and Data Security, Zhejiang University}
\country{China}
}
\email{xinghu@zju.edu.cn}

\author{Lingfeng Bao}
\affiliation{%
\institution{The State Key Laboratory of Blockchain and Data Security, Zhejiang University}
\country{China}
}
\email{lingfengbao@zju.edu.cn}

\author{Yu Liu}
\affiliation{%
\institution{The State Key Laboratory of Blockchain and Data Security, Zhejiang University}
\country{China}
}
\email{3200103741@zju.edu.cn}

\author{JianLing Sun}
\affiliation{%
\institution{The State Key Laboratory of Blockchain and Data Security, Zhejiang University}
\country{China}
}
\email{sunjl@zju.edu.cn}

\begin{abstract}
\input{abstract}

\end{abstract}




\maketitle

\section{INTRODUCTION}
\label{sec:intro}
\input{intro}

\section{Preliminary Study}
\label{sec:pre}
\input{pre}

\section{OUR APPROACH}
\label{sec:approach}
\input{approach}

\section{Empirical Evaluation}
\label{sec:eval}
\input{eval}

\section{Practical Application}
\label{sec:application}
\input{application}

\section{Related Work}
\label{sec:related}
\input{related}

\section{Threats to Validity}
\label{sec:threats}
\input{threat}

\section{Conclusion and Implications}
\label{sec:con}
\input{conclusion}

\begin{acks}
This research is supported by the Starry Night Science Fund of Zhejiang University Shanghai Institute for Advanced Study, Grant No. SN-ZJU-SIAS-001. 
This research is supported by the National Key Research and Development Program of China (No. 2021YFB2701102). 
This research is partially supported by the Shanghai Sailing Program (23YF1446900) and the National Science Foundation of China (No. 62202341, No.62372398, No.72342025, and U20A20173), and the Fundamental Research Funds for the Central Universities (No. 226-2022-00064). 
This research is partially supported by the Ningbo Natural Science Foundation (No. 2023J292). 
This research was also supported by the advanced computing resources provided by the Supercomputing Center of Hangzhou City University. 
The authors would like to thank the reviewers for their insightful and constructive feedback.
\end{acks}

\balance
\bibliographystyle{ACM-Reference-Format}
\bibliography{samples}


\end{document}

%% file: abstract.tex
Nowadays, more and more developers resort to Stack Overflow for solutions (e.g., code snippets) when they encounter technical problems. 
Although domain experts provide huge amounts of valuable solutions in Stack Overflow, these code snippets are often difficult to reuse directly. Developers have to digest the information within relevant posts and make necessary modifications, and the whole solution-seeking process can be time-consuming and tedious.  
To facilitate the reuse of Stack Overflow code snippets, Terragni et al. first explored transforming a code snippet in Stack Overflow into a well-formed method API (\textbf{A}pplication \textbf{P}rogram \textbf{I}nterface) by using a rule-based approach, named APIzator. 
The reported performance of their approach is promising, however, after our in-depth analysis of their experiment results, we find that (1) 92.5\% of APIs generated by APIzator are pointless and thus are difficult to use in practice. 
This is because the method name generated by APIzator (extracting \texttt{verb + object}) can rarely represent the method's functionality, which can hardly be claimed as meaningful/reusable APIs. 
(2) The authors manually summarized a number of rules to identify parameter variables and return statements for Java methods. 
These hand-crafted rules are extremely complex and sophisticated, and the manual rule design process is labor-intensive and error-prone. 
Moreover, since these rules are designed for Java, they can hardly be extended to other programming languages.

Inspired by the great potential of Large Language Models (LLMs) for solving complex coding tasks, in this paper, we propose a novel approach, named \toolname, to automatically perform APIzation for Stack Overflow code snippets. 
\toolname does not require additional model training or any manual crafting rules and can be easily deployed on personal computers without relying on other external tools. Specifically, \toolname guides the LLMs through well-designed prompts to generate well-formed APIs for given code snippets. 
To elicit knowledge and logical reasoning from LLMs, we used \textbf{c}hain-\textbf{o}f-\textbf{t}hought (CoT) reasoning and few-shot in-context learning, which can help the LLMs fully understand the APIzation task and solve it step by step in a manner similar to a developer. 
Our evaluations show that \toolname achieves a remarkable accuracy in identifying method parameters (65\%) and return statements (66\%) equivalent to human-generated ones, surpassing the current state-of-the-art approach, APIzator, by 15.0\% and 16.5\% respectively. 
Moreover, compared with APIzator, our user study demonstrates that \toolname exhibits superior performance in generating meaningful method names, even surpassing the human-level performance, and developers are more willing to use APIs generated by our approach, highlighting the applicability of our tool in practice. 
Finally, we successfully extend our framework to the Python dataset, achieving a comparable performance with Java, which verifies the generalizability of our tool.

%% file: intro.tex
\begin{figure}
	\centering
	\includegraphics[width = 0.95\linewidth]{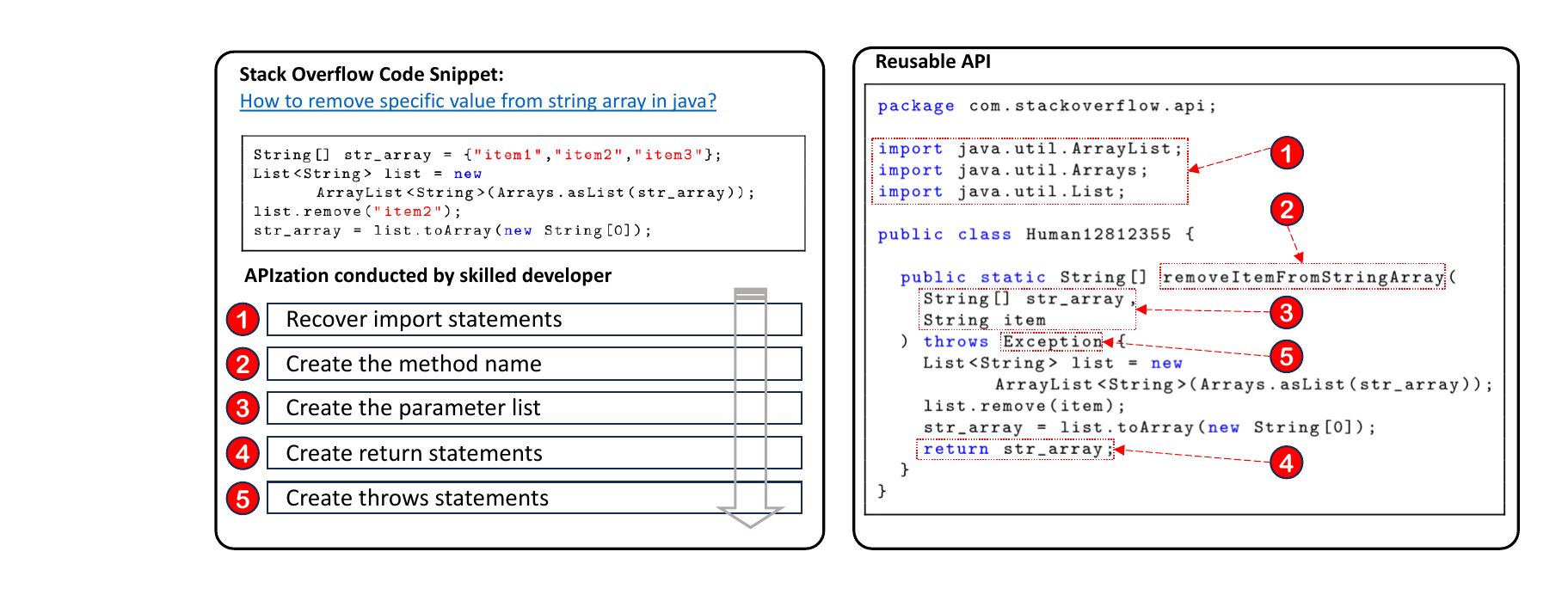}
	\caption{Human APIzation of A SO Code Snippet.}
	\label{fig:intro_demo1}
\end{figure}

Stack Overflow (SO), one of the most popular Software Q\&A (SQA) sites, plays an ever-increasing role in helping developers to solve their daily technical problems.  
Among the vast amount of knowledge shared on SO, code snippets hold particular significance as they are widely copied and pasted by developers, offering practical solutions to their daily programming problems. 
However, code snippets on SO are often unable to be used directly~\cite{terragni2016csnippex,subramanian2013making,yang2016query}. 
This is because SO code snippets are mainly written for illustrative purposes and do not focus on reusing purposes (such as missing import declarations)~\cite{nasehi2012makes}. Creating a reusable API for the SO code snippet requires substantial effort and is rather laborious and error-prone, even for accepted answers. 
Developers have to navigate through and digest the SO posts, and apply ad hoc modifications on code snippets for their own usage, such as carefully recovering missing import declarations, summarizing descriptive method names, extracting proper method parameters and replacing them in method bodies, inferring return statements and catching necessary exceptions. As reported by Terragni et al~\cite{terragni2021apization}, it takes more than four minutes for an experienced Java developer to build an API from a given SO code snippet. 


Consider the Java code snippet in Fig.~\ref{fig:intro_demo1} as an example. 
The Java code snippet is posted by an expert to solve the problem ''\textit{How to remove specific value from string array in java?}''. 
Making this code snippet off the shelf is difficult which requires various necessary steps. 
(1) Recovering missing import declarations. 
Missing variables, function definitions, or third-party dependencies can cause code to crash during compilation, we thus need to recover missing import statements (e.g., \texttt{import java.util.regex.ArrayList;}) at first. 
(2) Generating a meaningful method name that precisely describes what the method does. 
A good method name should be self-explained and intention-revealing, which can make the method much easier to understand, as well as to find and use. 
On the contrary, a meaningless method name can obscure the meaning of the code and waste developers' time for searching and reusing. 
Therefore, we need to create a descriptive and meaningful method name (i.e., \texttt{removeItemFromStringArray}) for the code snippet. 
(3) Identifying the method input parameters and abstracting them in code snippets. 
Extracting suitable variables as input parameters can provide good scalability for API reusing. 
For example, the variables \texttt{str\_array} and \texttt{item} are identified as input parameters passing to the API method. 
(4) Inferring the return statements. 
After identifying the input of the code snippet, we also need to infer and add output (i.e., \texttt{return str\_array}) for the API method. 
(5) Create throws statements. 
To safely use the API, it is necessary to declare exceptions that the method may throw. 
Finally, a skilled developer goes through the above steps and successfully outputs a reusable API as shown in Fig.~\ref{fig:intro_demo1}.

To automate the process of APIzation from SO code snippets, Terragni et al.~\cite{terragni2021apization} first proposed a tool, named APIzator, to convert Java code snippets to compliable APIs that developers can easily incorporate. 
The authors carefully designed a number of rules for Java to identify the method input parameters and return statements and employed the Part-Of-Speech (POS) Tagging technique to generate method names. 
As a result, APIzator takes the SO Java code snippet as input, after going through manually designed rules, APIzator outputs a compliable API as output. 
APIzator achieved a promising performance, for 81.5\% APIs generated by APIzator, either method parameters \textbf{or}  return statements are identical with ground truth. 
To understand the rationale for APIzator's good performance and its applicable scenario in practice, we conduct an in-depth analysis of their evaluation results. 
We find that: (1) \textbf{Most of the APIs generated by APIzator are hardly to be applied in practice.} 
According to our investigation, 90\% of these APIs are not applicable. 
This is because APIzator ignores the importance of meaningful names. 
APIzator generated method names by naively using the \texttt{verb} and its \texttt{object} from question titles, which are often imprecise and misleading. 
The inconsistency between method names and their implementations can confuse developers and even cause the introduction of bugs in the future~\cite{tan2007icomment, tan2011acomment, tan2012tcomment}.  
(2) \textbf{The rules designed for APIzator are too complex and specific, and can hardly generalize to other programming languages.} 
For example, APIzator heavily relies on tools (e.g., CSNIPPEX~\cite{terragni2016csnippex} and BAKER~\cite{subramanian2014live} for recovering missing variables and type declarations) and complex rules (e.g., using PATT-const to recognize hard-coded initializations), implementing APIzator and adjusting it to other programming languages require a substantial manual effort. 
Therefore, the key research question we ask in this work is: \textit{Can we design models to generate applicable APIs for SO code snippets and generalize to other programming languages easily?}

The recent success of ChatGPT~\cite{OpenAI-blog-2022-ChatGPT} based on GPT-3.5 demonstrates the remarkable ability of large language models (LLMs)~\cite{zhao2023survey,wang2023software,wang2021codet5,touvron2023llama,wang2022self,chowdhery2022palm} to comprehend human questions and assist in coding-related tasks.\footnote{https://platform.openai.com/docs/model-index-for-researchers} 
Inspired by the impressive capabilities of LLMs in code generation~\cite{fried2022incoder,feng2020codebert,clement2020pymt5,lu2021codexglue,jiang2023selfevolve,zheng2023codegeex}, in this work, we propose \toolname, a novel approach to automatically generate reusable and applicable APIs for SO code snippets. 
Notably, the underlying approach of \toolname is prompt engineering~\cite{liu2023pre,feng2023prompting}, i.e., prompting the tasks to generate desired output, which is extremely lightweight compared to rule-based methods with manually designed rules and ML-based methods with massive training data. 
\textbf{\toolname leverages few-shot learning~\cite{dong2022survey,brown2020language} and chain-of-thought reasoning~\cite{wei2022chain,kojima2022large} to elicit human knowledge and logical reasoning from LLMs to accomplish the APIzation task in a manner similar to a skilled developer.} 
Few-shot learning aims to solve the problem of how to train a model from a small number of examples. 
Regarding LLMs, few-shot learning is usually implemented using few-shot prompts~\cite{gao2020making,li2023codeie}.  
Specifically, a few task-specific examples are incorporated into the prompts to facilitate the model's understanding of desired input-output patterns for a given task. 
In this work, we employ few-shot prompts to help LLMs recognize the APIzation task. 
Moreover, we provide LLMs with detailed chain-of-thought reasoning from developers, allowing LLMs to refactor the code with the same thought process of developers (e.g., identifying method inputs/outputs, summarizing descriptive method names, handling exceptions and outputting a compliable API).

A comprehensive evaluation was conducted to evaluate the effectiveness of our tool. 
For a fair comparison, we reuse the evaluation set provided by Terragni et al.~\cite{terragni2021apization}, which contains 200 human-written APIs for SO code snippets. 
(1) Firstly, we assess the accuracy of identifying method parameters and return statements of \toolname, for 132 (65\%) and 130 (66\%) APIs, \toolname extracts equivalent method parameters and return statements with developers respectively. 
For 173 (86.5\%) APIs, either parameter list or return statements are equivalent. 
(2) APIzator is inapplicable because its generated method name is usually imprecise and misleading, we thus conduct a user study, wherein the quality of the method names generated by \toolname is compared against those produced by APIzator and even human developers. 
The user study shows that \toolname achieved human-level performance on generating high-quality method names and reusable APIs. 
(3) Instead of relying on complicated designed rules and requiring support from other tools, our framework is prompt-based and can easily apply to other programming languages. 
We verify the generalizability of our tool on Python, successfully creating 5,000 reusable APIs for SO Python code snippets. 
(4) We have implemented our tool as a Google Chrome extension to facilitate the developer's daily development. 
Our paper makes the following contributions: 
\begin{itemize}
    \item We thoroughly analyze the state-of-the-art tool, APIzator, and point out its limitations on APIzation tasks. 
    \item We propose an extremely lightweight and flexible approach, \toolname, that utilizes prompt engineering with few-shot learning and chain-of-thought reasoning to harness LLMs’ knowledge for creating reusable APIs. 
    The evaluations and user study show the superiority of our model over the state-of-the-art baseline. 
    Surprisingly, our approach achieves comparable or even better performance on this task than human developers. 
    \item We adapted our approach to Python for generalization, and we released a new dataset, which contains 6,023 reusable Java APIs and 5,000 reusable Python APIs generated from SO code snippets. These off-the-shelf APIs can speed up the searching and reusing of SO code snippets.  
    \item We have implemented our approach as a Chrome extension tool~\cite{Chrome-extension}, and released our replication package~\cite{replicate}, which can facilitate developers' daily development and inspire follow-up research.  
\end{itemize}

%% file: pre.tex

\subsection{\label{section2.1}Limitations of APIzator}

Terragni et al.~\cite{terragni2021apization} first introduced the task of APIzation. 
That is, for a given SO code snippet without method declaration, convert the code snippet into a functional and compilable API. 
To achieve this, the authors proposed a rule-based approach, namely APIzator, to complete this task automatically. 
They claim APIzator can generate ``\textbf{reusable}'' APIs and 81.5\% generated APIs are identical (either method parameters or return statements) to those produced by humans. 
However, this claim is rather shaky because an API can hardly be claimed as ``\textbf{reusable}'' if it only has correct method parameters or correct return statement, crafting a meaningful API method name is undeniably crucial for ensuring the true reusability of an API. 

\begin{figure}
	\centering
	\includegraphics[width = 0.95\linewidth]{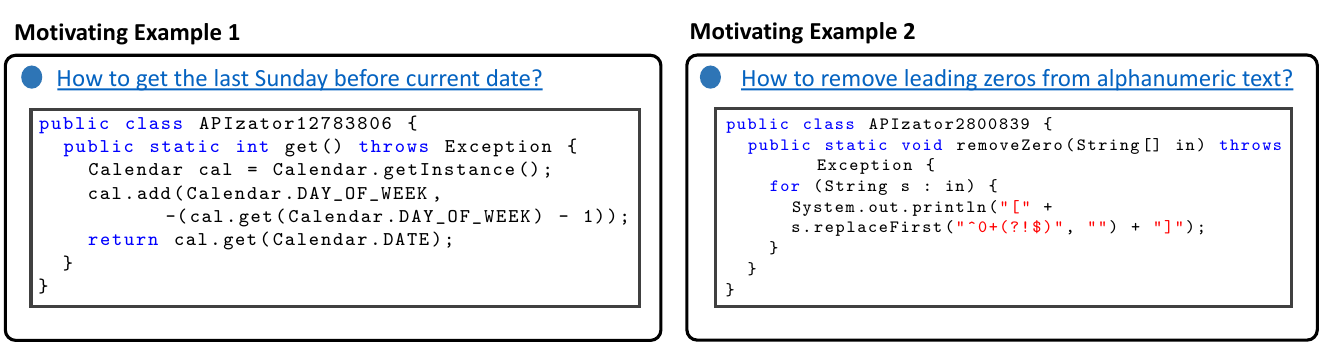}
	\caption{API Examples Generated by APIzator.}
    \vspace{-5pt}
	\label{fig:pre}
    \vspace{-5pt}
\end{figure}

In this preliminary study, we perform an in-depth analysis of their experimental results. 
Particularly, the first author of this paper manually examined their released evaluation set, which contains 200 API pairs (one generated by APIzator and the other one by human experts). 
It is worth mentioning that the preliminary study was conducted by the first author, which may introduce personal bias. A more comprehensive evaluation was conducted in Section~\ref{sec4.2}. Regarding the preliminary study, the examiner carefully reviewed each linked SO post to determine if the method names generated by APIzator were imprecise and/or misleading. As a result, 92.5\% (185) APIs' method names are not applicable (either imprecise or misleading) in practice. 
If we redefine the criteria of a "\textbf{reusable}" API to include not only identical method parameters and return statements but also meaningful method names, \textbf{the reusable API ratio of APIzator significantly drops from 81.5\% to 1.5\% (only 3 APIs meet this criteria)}. 
The reason for this phenomenon is that APIzator is a rule-based method, it generates method names by naively extracting \texttt{verb + object} from the question title, which is too simple to cover the complex scenarios in practice, resulting in a large number of misleading and meaningless method names. 
Another limitation of APIzator is that it is specifically designed for handling Java code snippets, its manually designed rules and third-party tools can hardly adapt to other programming languages. 

\subsection{Motivating Examples}
We provide two motivating examples from APIzator evaluation results as shown in Fig.~\ref{fig:pre}. 
In the first motivating example, APIzator generates a method named \texttt{get()} for the SO post ``\textit{How to get the last Sunday before current date?}'', the method name of this API is \textbf{unclear and meaningless}, developers who want to reuse this API have to carefully read through the method implementation to figure out what the method does. 
Moreover, the APIzator generated method names can even \textbf{mislead developers to misuse API and potentially lead to the introduction of bugs in future code.} 
For example, in motivating example 2, the APIzator naively extracted the \texttt{verb + object} from the question title (i.e., ``\textit{How to remove leading zeros from alphanumeric text}''), making a method name \texttt{removeZero()}  for this post. 
A developer can easily misuse this API by assuming this method removes all zeros from the input string, however, what this code snippet does is only removing prefixed zeros from a string.  
Overall, after manually examining Terragni et al.~\cite{terragni2021apization} released evaluation dataset, 197 out of 200 APIs can not be reused in practice due to their meaningless or misleading method names and limited method implementations, which motivates us to develop more advanced models for APIzation task.

%% file: approach.tex
\begin{figure}
	\centering
	\includegraphics[width = 0.90\linewidth]{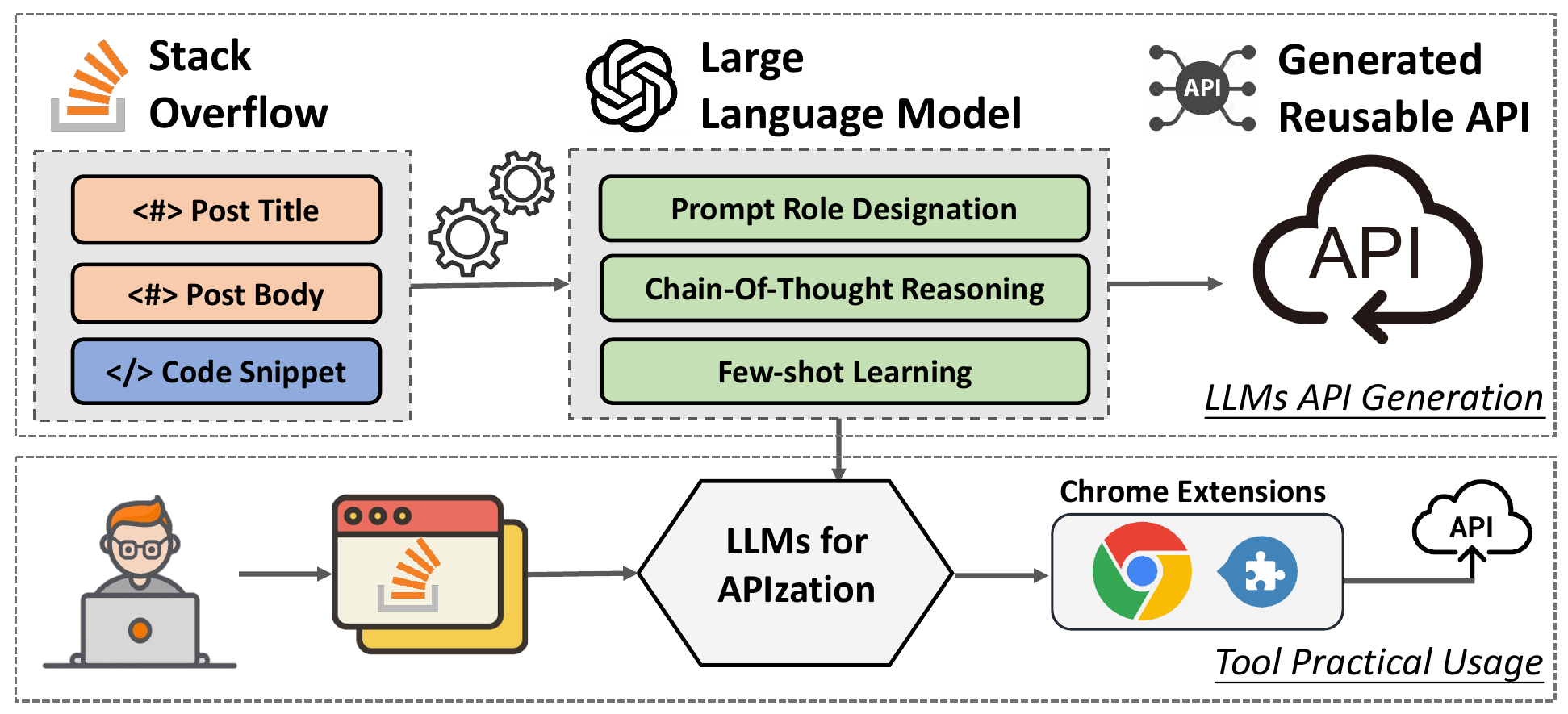}
    \vspace{-5pt}
	\caption{The Workflow of Our Approach.}
	\label{fig:workflow}
    \vspace{-5pt}
\end{figure}

In this section, we present a novel approach, namely \toolname, that leverages LLMs to transform SO code snippets into reusable APIs. 
The overall framework of our approach is illustrated in Fig.~\ref{fig:workflow}. 
As shown in Fig.~\ref{fig:workflow}, for a given SO code snippet, we first extract its associated question title and question body from the question post, 
then the question title, question body along code snippet are used to make our APIzation task-specific prompts. 
Since LLMs are not specifically designed to handle API generation, we employ prompt role designation, chain-of-thought reasoning, and few-shot learning strategy to guide LLMs to generate the desired output. 
Compared with APIzator, our approach is extremely lightweight without requiring any manually designed rules and/or massive training data. 

\subsection{Data Preparing}
Since our task targets SO code snippets, we downloaded the official SO data dump of March 2019 from StackExchange (the same data dump used by Terragni et al.~\cite{terragni2021apization}). 
The SO data dump contains timestamped information about the posts. 
Each post comprises a short question title, a detailed question body, corresponding answers, and multiple tags. 
The code snippets in SO provide solutions for practical problems, however, 
relying solely on these code fragments can pose difficulties in understanding their intended purpose. 
To enhance comprehension, we enrich the information provided to the LLMs by including the surrounding code context. 
In particular, for a given SO answer code snippet (enclosed by $\langle code \rangle$ tags), we retrieve its associated question title and question body, providing sufficient context for LLMs to capture the problem purpose and code semantics.

\subsection{Prompt Engineering}
The underlying approach of \toolname is prompt engineering, i.e., using natural language prompts to guide LLMs to complete specific tasks. In this work, we leverage \textbf{prompt role designation}, \textbf{chain-of-thought reasoning}, and \textbf{few-shot in-context
learning} to harness LLMs’ knowledge for automated API construction for SO code snippets.


\subsubsection{Prompt Role Designation.}
In prompt engineering, role designation is a method where LLMs are designated a role for solving a specific task. 
Assigning a role to LLMs provides it with the problem context that aids its understanding of the task context, and leads to more accurate and relevant responses. 
In this study, since we aim to convert code snippets (Java in this setting) into reusable APIs, we designate the role of LLMs to act as a skilled Java developer. 
After assigning the role to LLMs, we clearly inform the LLMs regarding our task as follows: ``\textit{Give you a context including a question title, a question post and an answer post, your task is to transform the Java code snippet within the answer post into Java method based on the context.}'' 
The prompt role designation explicitly indicates the input and output of the task, eliciting the programming knowledge of LLMs for completing this APIzation task.

\begin{table}
 \footnotesize
\caption{The Example of Prompt Engineering of Java APIzation.}
 \label{tab:prompt}
\centering
\begin{tabularx}{\textwidth}{m{0.1\textwidth}|X}
\toprule
\makecell{\textbf{Prompt}} & \makecell[Xt]{\textbf{Instantiation}} \\
\hline
\makecell{Role\\Designation} & \makecell[Xt]{Give you a context
including a question title, a question post and an answer post, your task is to transform the Java code
snippet within the answer post into Java method based on the context.}\\
\hline
\makecell[c]{Chain\\of\\Thought} & \makecell[Xt]{
To solve the problem, do the following:\\
Step 1 - Recover import statements based on the code snippet. If necessary, it can be none.\\
Step 2 - Define a public class Chatgpt that will be used to wrap the method.\\
Step 3 - Create ``public static'' modifier for the method.\\
Step 4 - Create the method name based on the context or the code snippet itself.\\
Step 5 - Infer parameter list based on the code snippet. If necessary, it can be none.\\
Step 6 - Infer return statements based on the code snippet. If necessary, it can be none.\\
Step 7 - Infer throws statements based on the code snippet. If necessary, it can be none.\\
Step 8 - Output the complete code based on the above results.\\
} \\
\hline
\makecell{Example\\input} & \makecell[Xt]{
Here are some examples:\\
Example one:\\
\textbf{Question title}:\\
How to convert \lstinline|int[]| into \lstinline|List<Integer>| in Java?\\
\textbf{Qustion post}:\\
How do I convert \lstinline|int[]| into \lstinline|List<Integer>| in Java? ..., I'll pick that one as the best to show the fact that this functionality is not part of Java.\\
\textbf{Answer post}: \\
There is no shortcut for converting from \lstinline|int[]| to \lstinline|List<Integer>| ..., you have to make a utility method. <the code snippet in this answer>\\
\textbf{Code snippet in the answer post}:\\
\lstinline|int[] ints = \{1, 2, 3\};|\\
\lstinline|List<Integer> intList = new ArrayList<Integer>(ints.length);|\\
\lstinline|for (int i : ints)\{|\\
\hspace{2em}\lstinline|intList.add(i);|\\
\lstinline|\}|\\
} \\
\hline
\makecell[t]{Example\\output} & \makecell[Xt]{
\textbf{Specific steps}: \\
Step 1: \lstinline|import java.util.ArrayList;import java.util.List;|\\
Step 2: \lstinline|public class Chatgpt \{\}|\\
Step 3: \lstinline|public static|\\
Step 4: \lstinline|convertIntArrayToList|\\
Step 5: \lstinline|(int[] arr)|\\
Step 6: \lstinline|return intList;|\\
Step 7: \lstinline|// None|\\
\textbf{Complete code:}\\ 
\lstinline|import java.util.ArrayList;|\\
\lstinline|import java.util.List;|\\
\lstinline|public class Chatgpt \{|\\
\hspace{2em}\lstinline|public static List<Integer> convertIntArrayToList(int[] arr) \{|\\
\hspace{4em}\lstinline|List<Integer> intList = new ArrayList<Integer>(arr.length);|\\
\hspace{4em}\lstinline|for (int i : ints)\{|\\
\hspace{6em}\lstinline|intList.add(i);|\\
\hspace{4em}\lstinline|\}|\\
\hspace{4em}\lstinline|return intList;|\\
\hspace{2em}\lstinline|\}|\\
\lstinline|\}|
} \\
\hline
\makecell{Test\\Input} & \makecell[Xt]{Now, give you the following context:
Question title: <Question title here>;
Question post: <Question post here>;
Answer post: <Answer post here>;
Code snippet in the answer post: <Code snippet here>} \\
\hline
\makecell{Format\\Constraints} & \makecell[Xt]{Please output the results in the following format:\\
\textbf{Specific steps:} <the results of step 1-7>\\
\textbf{Complete code:} <the result of step 8>} \\
\bottomrule
\end{tabularx}
\end{table}


\subsubsection{\label{section3.2.2}Chain-of-Thought Reasoning.}
Chain-of-thought reasoning is an important strategy for prompt engineering. 
It enables the LLMs to split a complex task into several relatively simple steps and generate a series of intermediate outputs that lead to a reasonable result. 

Our task of transforming a code snippet into a reusable API is a non-trivial task, it requires logical thinking to understand the SO post and a coherent series of intermediate steps to create APIs. 
In order to construct a reasonable CoT, based on previous research~\cite{feng2023prompting,wei2022chain}, we invited two developers with 12 years of Java programming experience for this task. 
Two developers were asked to manually transform 15 code snippets randomly extracted from the dataset released by APIzator (9,901 SO code snippets) into reusable APIs and write down their core steps as their chain-of-thought reasoning steps. 
Afterwards, the first author optimized their thinking process into an 8-step chain-of-thought reasoning by considering the following guidelines: (1) Write clear and specific instructions (e.g., “recover import statements based on the code snippet”); (2) Ask for a structured output (e.g., “please output the results in the following format”); (3) use delimiters to clearly indicate distinct parts (e.g., wrapping code with <>).
\textbf{
Finally, we designed an 8-step thinking process as chain-of-thought reasoning for LLMs as shown in Table~\ref{tab:prompt}, endowing LLMs to convert code snippets with the same thought process as skilled developers. 
}
Among them, \textit{Step4} - \textit{Step6} are the core steps of chain-of-thought reasoning, guiding LLMs to generate meaningful method names, identify input parameter lists and infer output return statements step-by-step. 
\textit{Step3} creates the default modifiers for the method. 
\textit{Step1} and \textit{Step7} are used to infer \texttt{import} and \texttt{throws} statements. 
\textit{Step2} and \textit{Step8} are used to facilitate subsequent data processing. 
Overall, this step-by-step thinking guides LLMs to generate expected reusable APIs. 


\subsubsection{Few-Shot Learning.}
With the increasing ability of LLMs, in-context learning has been widely adopted as zero-shot learning and few-shot learning. 
As LLMs are not specifically trained with SO code snippets, we adopt the few-shot learning strategy in this study. 
Few-shot learning is utilized to augment the context with a few examples of desired inputs and outputs, which helps the model elicit specific knowledge and abstractions needed to complete the task. 

To identify representative examples for few-shot learning, we first randomly extracted 100 examples from the dataset used in Section~\ref{section3.2.2}. 
We then selected our representative examples by the following criteria: (1) Since the CoT reasoning plays a vital role in guiding LLMs to perform the APIzation task step by step in a way similar to a developer, our first rule for choosing few-shot examples is to cover different steps (each comprising at least 7 steps) in the CoT reasoning process. 
(2) Considering the token limit of LLMs, we then removed the code samples that were too long or too short, only keeping moderate-length code snippets ranging from 3 to 10 lines. 
Following the filtering process, we identified 17 examples meeting the above criteria. 
Considering the input length limitations of LLMs mentioned in previous studies~\cite{feng2023prompting, wei2022chain}, after further discussing the quality and diversity of these examples with two developers in Section~\ref{section3.2.2}, we finally selected 5 representative examples, one of which is illustrated in Table~\ref{tab:prompt}.


\subsubsection{\label{sec3.2.4}Prompt construction.}
Our final prompt consists of six parts: prompt role designation, chain-of-thought reasoning, example input, example output, test input and format constraints. Each part plays a distinct role as follows: 
\begin{itemize}
    \item\textbf{Prompt Role Designation:} It offers a comprehensive overview of the SO code snippet APIzation task, setting the context for the subsequent steps.
    \item\textbf{Chain-of-Thought Reasoning:} It guides our model step by step, enabling the LLMs to solve the APIzation problem with the same thought process as developers. 
    \item\textbf{Example Input and Example Output:} 
    The examples further illustrate the task's requirements and show the desired output format, aiding the model in understanding the expectations more distinctly.
    \item\textbf{Test Input:} This part represents the problem that our model currently needs to solve, serving as a practical evaluation scenario.
    \item\textbf{Format Constraints:} They define the specific input form for the JAVA APIzation task and the expected output form of the LLM.
\end{itemize}
We directly connect these six parts as our final prompt, we have provided a complete prompt in our replication package~\cite{replicate}. 
After feeding the constructed prompts to the LLMs, the LLMs will output the result of APIzation which appears after the ``\textbf{Complete code}'' field as shown in Fig.~\ref{tab:prompt}. 
We use regular expressions to post-process the output of the LLMs and save the generated APIs in the ``Code2API\textbf{Id}.java'' file for evaluation, where ``\textbf{Id}'' refers to the answer Id of the corresponding SO answer post.

\subsection{The Implementation of the LLM}
For the LLMs, We chose the state-of-the-art GPT-3.5-turbo model,\footnote{https://platform.openai.com/docs/model-index-for-researchers} one of the best instruction-tuned LLMs~\cite{ouyang2022training,lou2023prompt} as our base model, which has been proven to have excellent abilities in tasks such as text summarization~\cite{yang2023exploring} and machine translation~\cite{hendy2023good}. 
To ensure the uniqueness of the experimental results, we set the temperature parameter to 0 during all experiments to make the output of LLMs consistent. 
Notably, during evaluation, only one code snippet's prompt (out of 200 code snippets) exceeded the maximum input token limit of GPT-3.5, which means the GPT-3.5-turbo model is sufficient to handle this APIzation task.

%% file: eval.tex
In this section, we conducted comprehensive experiments to evaluate the performance of our approach. 
Specifically, we aim to answer the following research questions:
\begin{itemize}
    \item \textit{RQ-1: How effective is our \toolname in identifying the method parameters and return statements compared with baselines?}
    \item \textit{RQ-2: How effective is our \toolname in generating meaningful method names \& are developers willing to use the APIs generated by our tool?}
    \item \textit{RQ-3: How effective do chain-of-thought reasoning and few-shot in-context learning contribute to the overall performance?}
    \item \textit{RQ-4: Can our \toolname easily generalize to other programming languages?} 
    \item \textit{RQ-5: How effective is our \toolname in generating compilable APIs?}
\end{itemize}


\subsection{RQ-1: Method Parameters and Return Statements Evaluation} 
\subsubsection{Experimental Setup.}
Terragni et al.~\cite{terragni2021apization} first introduced the task of APIzation (i.e., converting SO code snippets into reusable APIs) and proposed APIzator for this task. 
They established a benchmark comprising 200 APIzations performed by 20 developers.
Their evaluation dataset contains 200 pairs of APIs, one generated by the human developer and one generated by APIzator. 
In their research paper, they compared each pair to evaluate if APIzator can generate identical method parameters and return statements with human developers. 
For a fair comparison, we reused the evaluation dataset released by Terragni et al.~\cite{terragni2021apization}. 
If we extend their evaluation dataset with new evaluators, it may introduce extra bias from the labeling process. 
Particularly, for each code snippet in their evaluation dataset, we retrieve its associated question title, question body, and answer body to make our prompt (detailed in Section~\ref{sec:approach}).
Subsequently, we input each constructed prompt into LLMs to generate a reusable API. 
As a result, we also obtained 200 APIs generated by our \toolname. 
Therefore, We can make a pairwise comparison between our \toolname's generated API and human-generated API to estimate our approach effectiveness. 

\subsubsection{Evaluation Metrics.}
APIzator evaluated its effectiveness with respect to two aspects: the accuracy of identifying method parameters and the accuracy of identifying return statements. 
In this research question, we follow their experiment settings and first evaluate method parameters and return statements generated by our approach. 
In particular, we consider the following three evaluation metrics:
\begin{itemize}
    \item Equivalent Method Parameters: Given an API pair $\langle API_{1}, API_{2}\rangle$, let $P_{1}$ and $P_{2}$ denote the parameter list of $API_{1}$ and $API_{2}$ respectively.
    $P_{1}$ and $P_{2}$ are considered to be equivalent if they are both empty or contain identical parameters. 
    Two parameters are identical if they: (i) have the same type; (ii) refer to the same parameter in the method body (by manual inspection). 
    It is worth mentioning that we use a slightly different definition compared with Terragni et al.~\cite{terragni2021apization}, we do not require identical parameters to have same identifiers (i.e., variable names), since parameters with different identifiers can also be equivalent. 
    \item Equivalent Return Statements: We reuse APIzator's definition of the identical return statements~\cite{terragni2021apization} as equivalent return statements in this study. 
    That is, given an API pair $\langle API_{1}, API_{2}\rangle$, let $R_{1}$ and $R_{2}$ denote the return statement of $API_{1}$ and $API_{2}$ respectively. 
    $R_{1}$ and $R_{2}$ are considered as equivalent if they: (i) both have void as the return type; or (ii) have the same return type in the method header and have identical return statements in the method body. 
    \item Equivalent Method Implementation: Given an API pair $\langle API_{1}, API_{2}\rangle$, $M_{1}$ and $M_{2}$ denote the method implementation of $API_{1}$ and $API_{2}$ respectively. $M_{1}$ and $M_{2}$ are considered to be equivalent if: (i) $API_{1}$ and $API_{2}$ have equivalent method parameters; and (ii) $API_{1}$ and $API_{2}$ have equivalent return statements; (iii) the method body of $API_{1}$ and $API_{2}$ implement the same functionality. 
\end{itemize}


 


In this research question, we aim to evaluate our \toolname generated APIs and human-generated APIs in terms of the three aforementioned evaluation metrics: (1) the accuracy of equivalent method parameters (denoted as \texttt{P-Acc}); (2) the accuracy of equivalent return statements (denoted as \texttt{R-Acc}); (3) the accuracy of equivalent method implementations (denoted as \texttt{M-Acc}). 
(4) We use \texttt{PR-Acc} to denote the proportion of APIs with either equivalent parameter lists or equivalent return statements.

\subsubsection{Evaluation Results} 
The evaluation results of our \toolname and APIzator with respect to the above evaluation metrics are shown in Table~\ref{tab:RQ1_results}. 
The original performance of APIzator on these four evaluation metrics has been tested and reported (i.e., the \texttt{P-Acc} referred to their RQ-2, the \texttt{R-Acc} referred to their RQ-3, \texttt{M-Acc} referred to their RQ-1 and \texttt{PR-Acc} referred to their discussion) and we summarized them in Table~\ref{tab:RQ1_results} as APIzator (Ori). 
Since we make a slightly different definition on equivalent method parameters, we recalculate the \texttt{P-Acc}, \texttt{M-Acc},  and \texttt{PR-Acc} based on our current standards and denote as APIzator (Cur). 

\begin{table}
\footnotesize
  \caption{Performance Comparison of Different Approaches}
  \label{tab:RQ1_results}
  \centering
  \begin{tabular}{lcccc}
    \toprule
    \textbf{Approach} & \textbf{M-Acc} & \textbf{P-Acc} & \textbf{R-Acc} & \textbf{PR-Acc}\\
    \midrule
    APIzator (Ori) & 31.5\% & 56.5\% & 57.5\% & 81.5\% \\
    \midrule
    APIzator (Cur) & 32.5\% & 56.5\% & 57.5\% & 81.5\% \\
    Code2API & \textbf{43.5\%} & \textbf{65.0\%} & \textbf{66.0\%} & \textbf{86.5\%} \\
  \bottomrule
\end{tabular}
\end{table}

From Table \ref{tab:RQ1_results}, it can be seen that: 
\textbf{\toolname significantly outperforms APIzator in all evaluation metrics by a large margin}, achieving an \texttt{M-Acc} of 43.5\%, a \texttt{P-Acc} of 65.0\%, an \texttt{R-Acc} of 66.0\%, and a \texttt{PR-Acc} of 86.5\%, These are respectively 11.5\%, 8.5\%, 8.5\%, and 5\% higher than the corresponding metrics of APIzator. 
In other words, our approach is on average 15.0\% and 16.5\% more accurate than the state-of-the-art tool, APIzator, in identifying equivalent method parameters and equivalent return statements. 
We attribute this to the following reasons: 
(i) Compared with APIzator heavily relies on manually crafted rules, LLMs have their own ability to perform logical inference and deductive reasoning; 
(ii) We use chain-of-thought reasoning in our prompt engineering, endowing LLMs to perform APIzation with the same thought process of developers; 
(iii) We also use few-shot in-context learning in our prompt engineering, guiding LLMs to infer correct outputs step-by-step. 
\subsubsection{Manual Analysis.}
\begin{figure}[h]
	\centering
	\includegraphics[width = 0.95\linewidth]{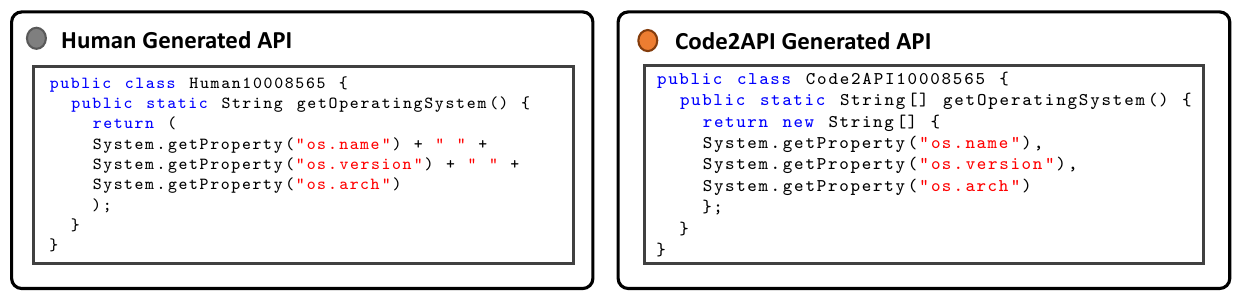}
    \vspace{-10pt}
	\caption{An Example of Reasonable APIs generated by Code2API.}
	\label{fig:reasonalbe}
    \vspace{-10pt}
\end{figure}

\begin{figure}[h]
	\centering
	\includegraphics[width = 0.95\linewidth]{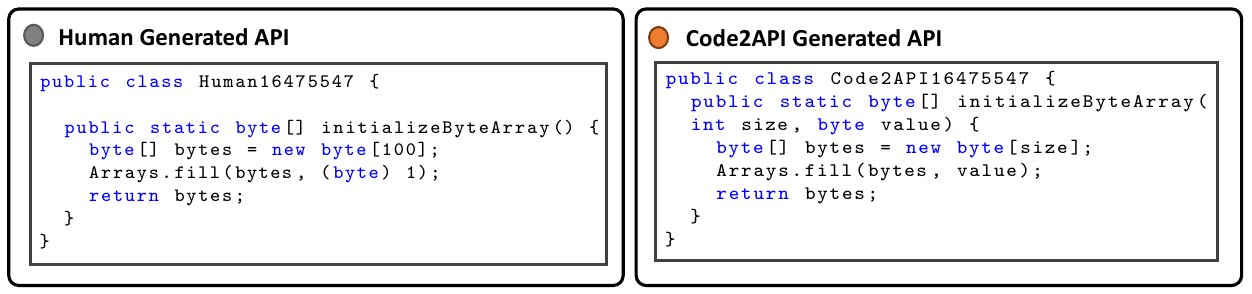}
    \vspace{-10pt}
	\caption{An Example of Better APIs Generated by \toolname.}
	\label{fig:better}
    \vspace{-10pt}
\end{figure}

As can be seen from Table~\ref{tab:RQ1_results}, there are a number of APIs generated by our approach that are not equivalent to humans'. 
To explore the reason why our approach fails, we analyzed all error cases where \toolname fails to generate "\textit{correct method parameters}" and/or "\textit{correct return statements}", and categorized them into the following three types: (1) Reasonable or superior to human-generated APIs; (2) Missing necessary method parameters or return statements; (3) Others. 
We detailed the analysis of these three types of failed cases as follows: 




Regarding type1 failed cases, 30 failed cases of method parameters and 32 cases of return statements belong to this category. 
In RQ-1, we only consider exact matches with human-generated APIs as "correct", however, a code snippet has various forms of reusable APIs. 
If these type1 failed cases are recounted, the performance of \toolname can be further improved by 15\% and 16\% respectively. 
For example,
\textbf{a common failed situation is that our approach-generated APIs are different from humans' but reasonable}. 
Different human developers may create different reusable APIs depending on their coding preferences. In other words, for a given SO code snippet, there is more than one "correct answer" in terms of its reusable API. 
An example of this situation is shown in Fig.~\ref{fig:reasonalbe}. 
As can be seen, for the SO post, ``\textit{How to get operating system in Java?}'', the human-generated API is shown on the left while \toolname generated API is shown on the right. 
It is clear that \toolname generated the same method name and method body as humans, the only difference between these two APIs is the return type. 
The human developer returns the operating system information as a \texttt{String} type by concatenating \texttt{os.name}, \texttt{os.version} and \texttt{os.arch} together, while \toolname returns the information as an \texttt{Array} of strings. 
Even the return types are different, both API implementations are reasonable and can be considered as reusable. 
\textbf{Another common failed situation is that our approach-generated APIs are better than the ones written by humans}. 
An example of this situation is shown in Fig.~\ref{fig:better}. 
The code snippet was extracted from the SO post ``\textit{How to initailize byte array of 100 bytes in java with all 0's}'', the skilled developer refactored this code snippet into a method named \texttt{initializeByteArray()}. 
However, this API is still difficult to reuse in practice because the API is specifically written to solve the above post, this API is hard to scale to other users' requests by fixing the size (i.e., 100) and value (i.e., 1) of the byte array. 
Compared with human-written APIs, our approach-generated APIs are more scalable and easy to use, regarding the above SO code snippet, \toolname successfully inferred the \texttt{size} and \texttt{value} as method parameters for initializing a byte array, allowing the APIs can be reused by developers with diverse requests.

Regarding type2 failed cases, 25 method parameters failed cases and 31 method return statement failed cases belong to this category. 
\toolname failed to identify necessary parameters and/or return statements for this category, because some parameters are too subtle (e.g., list index) and thus difficult to be detected by LLMs. 
Regarding type3 failed cases, 15 method parameters failed cases and 5 return statement failed cases fall into this type. 
LLMs are not perfect, they can return the wrong parameter types or irrelevant return statements mismatching code context or method design. 
Employing or fine-tuning LLMs for code (e.g., CodeLlama~\cite{roziere2023code}) may improve our model’s performance on this task, which is an interesting research direction for our future work.


\subsection{\label{sec4.2}RQ-2: Method Name Quality and API Preference Evaluation}
\subsubsection{Experimental Setup.}
As mentioned in Section~\ref{sec:pre}, one of the key limitations of APIzator is it generates a large number of meaningless and misleading method names. 
They ignored the importance of a descriptive method name and did not evaluate method name quality in their study. 
Therefore, in this research question, we aim to evaluate the method name quality generated by our approach and benchmarks (including APIzator and human-generated APIs). 
Since it is difficult for automatic evaluation to judge the semantic correctness of a method name, we conducted a user study for this research question. 
Terragni et al.~\cite{terragni2021apization} invited 20 participants with 9.8 years of Java programming experience on average to make the ground-truth dataset, to reduce the bias of human evaluators, we conducted our human evaluation with a comparable experiment scale (18 participants) and evaluators with similar Java programming experience (10.9 years on average). The volunteers in our user study include 10 PhDs, 2 Postdoc researchers, and 6 Java developers.
All of whom are not co-authors, major in computer science and/or software engineering and have 10.9 years of experience in Java programming on average (min 4, median 10, and max 15). 
We reuse the evaluation dataset in RQ-1, which contains 200 API triplets, $\langle API_{H}, API_{A}, API_{C} \rangle$, representing the APIs generated by human, APIzator, and our \toolname respectively. 
We then divided 200 API triplets into six sub-datasets (each sub-dataset contains 33-34 API triplets). 
After that, each sub-dataset was evaluated by three volunteers independently. 
In particular, each volunteer was given a SO post (including the question title and a link to the post) and three APIs, he/she needed to do the evaluation in terms of two aspects: method name quality and overall API preference (detailed in Section~\ref{rq2_metrics}). 
It is worth mentioning that participants did not know which API was generated by which method.

\subsubsection{\label{rq2_metrics}Evaluation Metrics.} 
We use \textit{method name expressiveness} to evaluate the quality of a method name and \textit{the willingness to use} to evaluate the overall API usage preference respectively. 

\noindent\textbf{Method Name Expressiveness}: 
Method name expressiveness refers to the matching degree between the method name and the method implementations. 
In our user study, each volunteer was asked to rate the expressiveness of a method name on a scale between 1 and 4 by reading the method name, method implementation, and its associated posts. 
Score 1 indicates the method name and the method body implementations are irrelevant. 
Score 2 indicates the method name and the method body are of low relevance. 
Score 3 indicates the method name and the method body are of high relevance. 
Score 4 indicates the method name and the method body match exactly and the method name can fully express the intention of the method.

\noindent\textbf{Willingness to Use}:
Willingness to use measures the best API a user prefers when they perform daily software development. 
Willingness to use justifies how likely the generated APIs can elicit further practical usage in software development. 
In our user study, for a given API triplet $\langle API_{H}, API_{A}, API_{C} \rangle$, after each participant rates method name expressiveness for each API candidate, we ask them to choose the best API from the three candidates by their own developing experience. 
The evaluators were blinded as to which API was generated by which method. 
Since different users have their own preferences for choosing the best APIs, the final results are determined by the majority of voting. 
When the best API selected by three evaluators was inconsistent, the first author played the role of a mediator to discuss with the corresponding three evaluators to reach a consensus. 
Among human evaluation processes for selecting the best API, only two cases suffered from this situation, the influence of the first author is rather limited.

\subsubsection{Evaluation Results.} 
Finally, we collected 600 groups of human evaluation results from 18 evaluators.  
Each group contains three method name expressiveness scores and the best API among the three API candidates. 
Since each API was evaluated by three different evaluators, we combined all six sub-datasets human evaluation results and calculated the Cohen's Kappa~\cite{cohen1968weighted} coefficient between the three groups of user ratings, which were 0.67, 0.7, and 0.71 respectively, all greater than 0.6. 
This indicates a substantial agreement among the different groups of user ratings~\cite{landis1977measurement}. 
All evaluation results are demonstrated in Table~\ref{tab:RQ2_R}.

\begin{table}
\footnotesize
\caption{The Method Name Expressiveness Scores and Willingness to Use}
\label{tab:RQ2_R}
\centering
\begin{tabular}{lllllll|r}
\toprule
\textbf{MNE SCORE} & 1 & 2 & 3 & 4 & Avg & P-value & \textbf{WILLINGNESS}\\
\hline
Human & 25 (4.2\%) & 77 (12.8\%) & 138 (23.0\%) & 360 (60.0\%) & 3.39 & 1.27e-8 & 96 (48.0\%)\\
APIzator & 136 (22.7\%) & 248 (41.3\%) & 156 (26.0\%) & 60 (10.0\%) & 2.23 & 1.85e-122 & 3 (1.5\%)\\
\toolname & 14 (2.3\%) & 29 (4.8\%) & 111 (18.5\%) & 446 (74.3\%) & 3.65 & - & 101 (50.5\%) \\
\bottomrule
\end{tabular}
\end{table}

From Table~\ref{tab:RQ2_R}, we can observe that: (1) \textbf{APIzator achieved the worst performance regarding method name expressiveness.}
Only 10.0\% method names generated by APIzator get a score of 4, and 64\% method names are rated by developers as low-quality (score 1 and score2), which means more than half of the method names generated by APIzator are irrelevant and are difficult to reuse directly. 
(2) \textbf{Users mainly rate APIzator's method names as Score 3 or Score 2 (i.e., 67.3\%),} this is reasonable because APIzator extracted the \texttt{verb + object} from the question title as method names, which has a relationship with the code snippets. 
However, this Part-Of-Speech tagging technique is too simple to cover complex scenarios in practice, 22.7\% method names are irrelevant with the method implementation. 
(3) \textbf{Our approach outperforms APIzator by a large margin regarding the method expressiveness, more surprisingly, the method names generated by our tool are even better than the human-written method names.}
74.3\% method names generated by our approach obtain a score of 4 regarding method name expressiveness and 18.5\% method names obtain a score of 3. 
In other words, our model rarely outputs low-quality or irrelevant method names (i.e., only 7.1\%). 
We attribute the promising performance of \toolname on generating method names to the following reasons: (a) the LLMs' great potential for natural language understanding and logical reasoning; (b) we use chain-of-thought reasoning to endow LLMs to perform APIzation with the same thought process of developers; (c) we use few-shot in-context learning to guide LLMs to infer correct outputs step-by-step; (d) we provide sufficient context (e.g., question title, question body and answer body) to LLMs for inference. 
(4) We also conducted Mann-Whitney U rank tests~\cite{shier2004statistics} to calculate the p-values between our approach and each of the baselines. 
The p-values are substantially less than 0.01, which shows \textbf{the method name expressiveness scores of our model are significantly better to APIzator and Human}. 
(5) The last column of Table~\ref{tab:RQ2_R} shows users' preferences (i.e., willingness to use) when picking the best API from three candidates. 
As can be seen, 50.5\% of user selections chose our approach-generated APIs as their first choice, while 48.0\% chose human-generated APIs as best, and only 1.5\% selected the APIzator-generated APIs. 
The best API evaluation results are consistent with the method name quality results, demonstrating that, \textbf{for this APIzation task, our proposed approach has achieved a comparable or even better performance than skilled human developers.} 

\subsubsection{Manual Analysis.}
\begin{figure}[h]
	\centering
	\includegraphics[width = 1.0\linewidth]{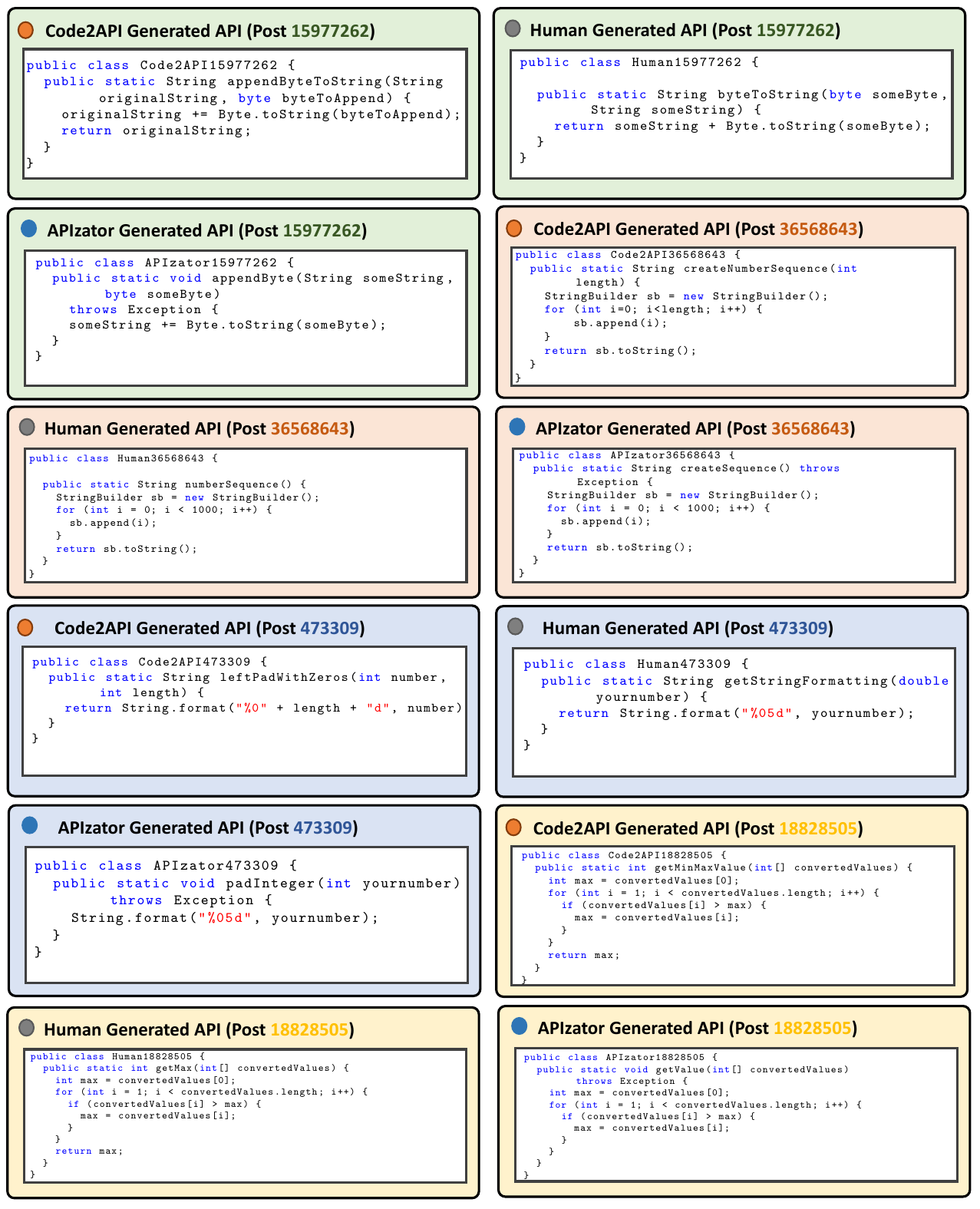}
	\caption{API Examples Generated by Different Methods}
	\label{fig:RQ2_example}
\end{figure}

To investigate the reasons why our approach achieves remarkable performance in method name expressiveness and willingness to use, we carefully investigated a number of API triples $\langle API_{H}, API_{A}, API_{C} \rangle$ where $API_{C}$ is the best.
The manually examined examples are shown in Fig.~\ref{fig:RQ2_example}. 
From these examples, we can see that: 
(1) Regarding the method name expressiveness, the advantage of our approach is obvious. 
As shown in the first group of API triplets (colored in green), given the code snippet from the post ``\textit{How to append a byte to a string in Java?}'', the method name generated by \toolname (i.e., \texttt{appendByteToString}) is more clear and precise than the name created by humans (i.e., \texttt{byteToString}) and APIzator (i.e., \texttt{appendByte}). 
Moreover, the method parameters extracted by \toolname are more meaningful and descriptive (e.g., \texttt{byteToAppend}) than human-summarized ones (e.g., \texttt{someByte}). 
This suggests that summarizing meaningful names is a non-trivial task for humans, which can be time-consuming and error-prone. 
\toolname can assist developers effectively. 
(2) Regarding the willingness to use metric, users prefer to adopt our approach generated API first. 
As shown in the second group (colored in orange), the code snippet comes from the post ``\textit{How to create a sequence of numbers in java}'', only \toolname inferred the \texttt{length} as a method parameter, allowing users to create arbitrary length number sequence. 
Another example is shown in group 3 (colored in blue), similarly, only APIzator allows users to add any number of zeros to the left through the parameter \texttt{length}. 
In addition, the \toolname generated method name \texttt{leftPadWithZeros} is self-explained, much more clear than \texttt{getStringFormatting} (generated by humans) and \texttt{padInteger} (generated by APIzator). 
It is no surprising that all users prefer to adopt our approach generated APIs for this case. 

\toolname rarely generates irrelevant APIs for SO code snippets, only 9 cases are rated by developers as low-quality (i.e., scored below 3). 
After manual investigation, these low-quality cases are mainly caused by three reasons: (1) Inconsistency between question title and code functionality (5 cases). The fourth group (colored in yellow) of Fig.~\ref{fig:RQ2_example} shows such an example. 
The code snippet is to get the maximum value of an array, however, APIzator generated \texttt{getMinMaxValue} method name for this code, which is inconsistent with the method implementations. 
The reason for this situation is that \toolname not only considers the semantics of the code snippet, but also the post question title ``\textit{how to get the minimum, maximum value of an array?}'', the inconsistency between the post and its code solution misguided our \toolname to generate incompatible method names. 
Since \toolname relies on post context for generation, these noise data should be detected and removed for our model in future work. 
(2) Exceeding max token limit (1 case). One case failed to get responses from Code2API because it exceeded the maximum number of input tokens of GPT-3.5-turbo (e.g., 4,096). To fundamentally solve this problem, it is necessary to increase the model's maximum input token limit. 
(3) Missing important information in method names (3 cases). Three method names generated by \toolname are considered overly abstract and/or drop valuable information. One possible solution is extracting important knowledge (such as keywords or terms) from posts and code for prompt construction.

\subsection{RQ-3: Ablation Analysis}
\subsubsection{Experimental Setup.} 
There are two key strategies we use in constructing our prompts, i.e., Chain-of-Thought (CoT) reasoning and few-shot learning. 
In this research question, we conduct an ablation analysis to verify their effectiveness one by one. 
Specifically, we compare \toolname with three of its incomplete versions: (i) Drop few-shot: we do not use few-shot learning when constructing prompts; (ii) Drop CoT: we do not use Chain-of-Thought when constructing prompts; (iii) Drop both: we drop both few-shot learning and CoT when constructing prompts. 
We reuse the evaluation metrics described in RQ-1 for comparison purposes. 

\subsubsection{Evaluation Results.}
The evaluation results are shown in Table~\ref{tab:rq3_performance}. It can be seen that: (1) No matter which prompt strategy we dropped, it hurts the overall performance of our model. This verifies the \textbf{importance and necessity of adding few-shot learning and chain-of-thought reasoning into our prompt engineering}. 
(2) Compared with Dop few-shot, Drop CoT has a better performance on \texttt{P-Acc}. 
While Drop few-shot has a better performance on \texttt{R-Acc} as compared with Drop CoT. 
This signals that \textbf{few-shot learning and chain-of-thought reasoning can complement and enhance the performance of each other} in generating reusable APIs. 
(3) After removing both, the performance of \toolname drops sharply, which is much lower than the original APIzator. 
This suggests that solely using LLMs is unable to solve the APIzation problem, \textbf{designing prompts that suit the specific task is the key of applying LLMs}. 
This further confirms the effectiveness of few-shot learning and CoT reasoning to elicit human knowledge and logical reasoning from LLMs to accomplish APIzation in a manner similar to a developer.

\renewcommand{\arraystretch}{1.05}
\begin{table}
    \footnotesize
    \tabcolsep=0.12cm
	\centering
	\caption{Performance Comparison of Different Prompts.}
	\label{tab:rq3_performance}
	\begin{tabular}{lcccc} 
	    \toprule
            \bf{Prompt} & \bf{M-Acc} & \bf{P-Acc} & \bf{R-Acc} & \bf{PR-Acc}\\
	      \midrule




            w/o few-shot & 22.5\% & 36.0\% & 55.0\% & 68.5\% \\

            w/o CoT & 30.5\% & 41.0\% & 53.5\% & 64.0\%\\

            w/o both & 6.5\% & 8.0\% & 10.5\% & 12.0\% \\
            \midrule
            \toolname & \textbf{43.5\%} & \textbf{65.0\%} & \textbf{66.0\%} & \textbf{86.5\%} \\
            \bottomrule
	\end{tabular}
\end{table}

\subsection{RQ-4: Generalization Study} 

\subsubsection{Experimental Setup}
As mentioned in Introduction~\ref{sec:intro}, another limitation of APIzator is that it is heavily designed and too complex to generalize to other programming languages. 
Compared with APIzator, our proposed \toolname is based on prompt engineering, which is extremely lightweight and flexible. 
In this research question, we aim to investigate the generalizability of our approach. 
In particular, we want to evaluate whether our approach can be easily adapted to Python and achieve a comparable performance with Java. 

To perform this generalization study, we collected 5000 appropriate code snippets from SO Python posts following the data collection process of APIzator (e.g., choosing ``how to'' questions, accepted answers with a score of at least 2, posts with exactly one code snippet, SO pages views in top 20,000). 
We then selected the top 100 Python code snippets (based on SO page views) and manually crafted APIs for these code snippets as our ground truth.  
After that, we slightly modified our chain-of-thought reasoning and few-shot learning from handling Java code snippets to Python code snippets (e.g., python functions do not need to be enclosed within a class, details can be found in our replication package~\cite{replicate}).  
Finally, we apply our framework to the Top 100 Python code snippets and generate reusable APIs for these SO posts. 
We used the same evaluation metrics in RQ-1 and compared its performance with Java evaluation results. 

\subsubsection{Evaluation Results.}
The evaluation results of our framework on the Python dataset are shown in Table~\ref{tab:RQ4_results}. 
We provide Java performance here for comparison. 
From the table, it can be seen that after translating prompts from Java to Python, our framework achieves a better performance regarding all evaluation metrics. 
A possible reason is that Python variables don't need to declare the data type explicitly, it is thus easier for \toolname to infer Python method parameters and return statements. 
\textbf{The consistently good performance of Code2API on Python dataset confirms that our framework can be easily generalized to other programming languages without any performance loss.} 


\begin{table}
  \footnotesize
  \caption{Performance Comparison of Different Programming Language}
  \label{tab:RQ4_results}
  \centering
  \begin{tabular}{lcccc}
    \toprule
    \textbf{Language} & \textbf{M-Acc} & \textbf{P-Acc} & \textbf{R-Acc} & \textbf{PR-Acc}\\
    \midrule
    Java & 43.5\% & 65.0\% & 66.0\% & 86.5\% \\
    Python & 57.0\% & 69.0\% & 80.0\% & 92.0\% \\
  \bottomrule
\end{tabular}
\end{table}


\subsection{RQ-5: Compilation Rate Analysis}
\label{sec:compilation}
\subsubsection{Experimental Setup}
Generating compilable APIs is crucial for developers to reuse APIs in their daily development. 
In this research question, we aim to investigate the compilation ratio of APIs generated by our approach and other baselines on the same evaluation dataset of RQ-1. 
In particular, we compare \toolname with the following three approaches: (1) Original Code Snippet: for a given code snippet, we directly run it to see if it is compilable; (2) PostFinder~\cite{rubei2020postfinder}: the component of PostFinder can wrap a code snippet into a compilable code, which can be regarded as a baseline for this research question. 
For a given code snippet, we wrap it with PostFinder and test if it is compilable; 
(3) APIzator: for comparison, we calculate the compilation ratio of APIs generated by APIzator. 

\renewcommand{\arraystretch}{1.05}
\begin{table}[H]
    \footnotesize
    \tabcolsep=0.12cm
	\centering
	\caption{Compilation Rate Analysis}
	\label{tab:diss}
	\begin{tabular}{lcccc} 
	    \toprule
            \bf{Method} & Original Code Snippet & PostFinder & APIzator & \toolname\\
            \midrule
            \bf{Compilation Rate} & 14.5\% & 46.5\% & \textbf{99.5\%} & 95\% \\
            \bottomrule
	\end{tabular}
\end{table}
\subsubsection{Experimental Results}
The experimental results are shown in Table~\ref{tab:diss}, it can be seen that: 
(1) The compilation rate of original code snippets is only 14.5\%, which further confirms that code snippets on SO are not able to be reused directly. 
(2) Compared with APIzator and \toolname, PostFinder achieves the worst performance due to its simple fixing strategy of using text matching. 
(3) APIzator achieves the highest compilation rate (e.g., 99.5\%) because of its use of existing tools (CSNIPPEX/BAKER) for fixing compilation errors. 
However, the meaningless method names greatly hinder the usage of APIzator’s APIs. 
(4) 95\% APIs generated by \toolname can be successfully compiled. 
It’s worth mentioning that if we feed the compilation error message to LLMs for regeneration, \toolname can ultimately resolve all the compilation errors, further verifying the practical and potential usage of our tool. 
In general, \toolname does not rely on specific compilation error fixing tools (like CSNIPPEX/BAKER) and is highly effective for generating compilable and reusable APIs. 
Moreover, it has the potential to solve different types of compilation errors by interacting with LLMs for multiple rounds.



%% file: application.tex
Regarding the time efficiency, for the 200 code snippets of the Java evaluation dataset, the average time cost of running our approach is 10 seconds per code snippet, which is comparable to APIzator execution time (8s for each code snippet). 
As reported by Terragni et al.~\cite{terragni2021apization}, developers need 4min and 22s on average to perform a single APIzation. 
Considering the remarkable performance of our \toolname in generating reusable APIs is comparable with human developers and the generation time cost with our approach is efficient, we have implemented our \toolname as a Chrome extension~\cite{Chrome-extension}, which can facilitate developers in using our approach for daily development when reusing SO code snippets and inspire follow up research for this.\\ 

\begin{figure}[h]
	\centering
	\includegraphics[width = 0.95\linewidth]{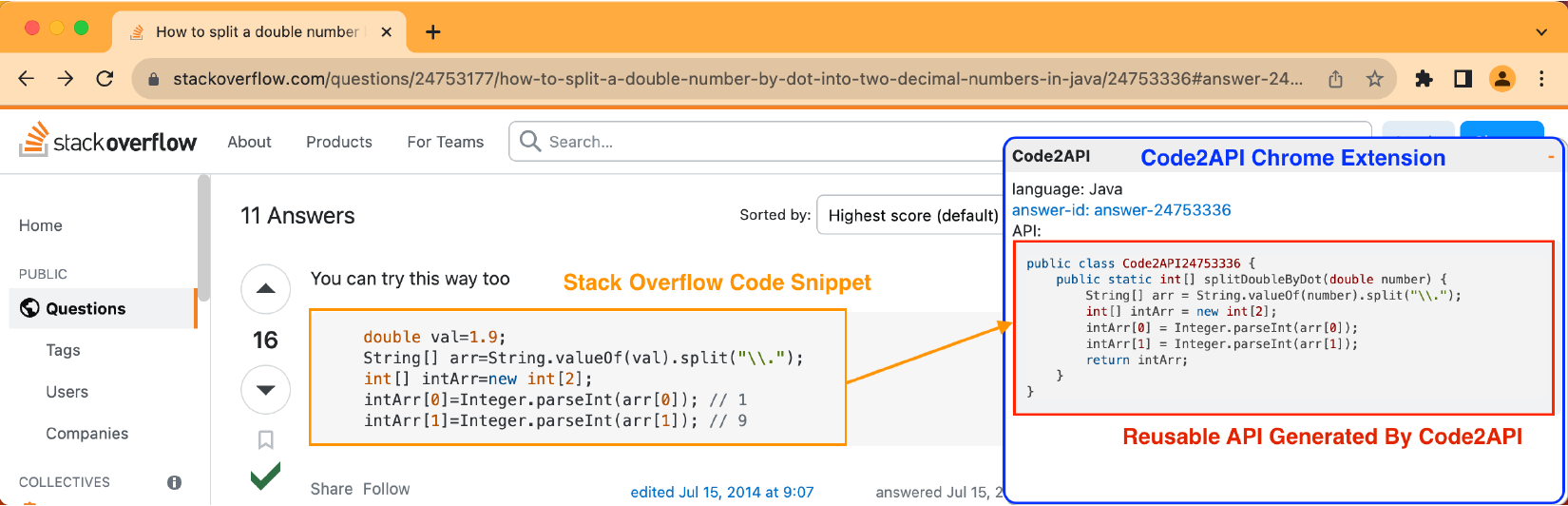}
	\caption{The Overview of Our Chrome Extension}
	\label{fig:extension}
    \vspace{-7pt}
\end{figure}

\noindent\textbf{Tool Implementation.}
We developed a Chrome extension, \toolname, to help developers automatically convert a SO code snippet into a reusable API. \toolname consists of two components: a frontend part (running on the user's browser) responsible for fetching the code snippet on SO and displaying the generated API, and a backend part (using GPT-3.5-turbo) responsible for generating the reusable API with our prompt engineering. 
Once the SO page is loaded, the frontend will prompt the user to click on the code snippet they are interested in. It will then call the backend model to construct the prompt and use GPT-3.5-turbo to generate the reusable API. Finally, the frontend alters the page to present the result to the user.
The tool overview is shown in Fig.~\ref{fig:extension}. 

\noindent\textbf{Practical Value.}
For experienced developers, our tool can greatly improve their efficiency in reusing code snippets (10 seconds per API, 26 times faster than humans). 
It can also serve as an API design tool for novice developers, who can learn recommended practices by examining the generated outputs. 
Moreover, \toolname has the potential to benefit development teams by standardizing the way code is reused from SO, fostering better collaboration and code quality. 
We will continue to optimize our tool and extend it to other programming languages in the future.

%% file: related.tex
\subsection{Code Generation}
Code generation~\cite{fried2022incoder,feng2020codebert, clement2020pymt5, lu2021codexglue, jiang2023selfevolve, zheng2023codegeex, yan2023closer} is a significant topic in software engineering research, 
the goal of this task is to transform a given natural language description into its corresponding code implementations. 
With the emergence of the code pre-training model CodeBERT~\cite{feng2020codebert}, more and more work attempts to solve the code generation task using the large pre-trained models. 
For example, Clement et al.~\cite{clement2020pymt5} proposed PYMT5 to complete the task of converting Python methods and method document strings. 
Lu et al.~\cite{lu2021codexglue} proposed CodeGPT to generate member functions and member variables in Java classes. 
Unlike the existing code generation research, the input of APIzation task is the SO code snippet, while the expected output is the reusable APIs of the code snippet. 
Our model considers the code snippet and its context (e.g., post title, post body) for generating reusable APIs for developers. 

\subsection{SO Mining and API Calls Generation}
Stack Overflow, one of the world's largest programmer Q\&A communities, has accumulated a vast number of technical questions and answers. 
Currently, many studies have been conducted on SO, including post recommendation~\cite{ponzanelli2014mining,rubei2020postfinder,gao2020technical, gao2023know}, query reformulation~\cite{cao2021automated, gao2020generating, gao2021code2que}, and content quality analysis~\cite{zhang2019empirical}. 
Specifically, Prompter~\cite{ponzanelli2014mining} and PostFinder~\cite{rubei2020postfinder} utilize the code context within the IDE to search for relevant posts on SO. 
BIKER~\cite{huang2018api} used the word embedding techniques to bridge the gap between natural language description and API documentation. 
FOCUS~\cite{nguyen2019focus} is a recommender system to provide developers with suitable API function calls and code snippets. 
DeepAns~\cite{gao2020technical} uses weakly supervised learning to recommend the best answer for a given SO question. 
Cao et al.~\cite{cao2021automated} implemented automatic query reformulation using query log data from SO. 
Zhang et al.~\cite{zhang2019empirical} conducted an empirical study confirming that many answers on SO are obsolete. 
Different from the above studies, \toolname aims at transforming SO code snippets into reusable APIs by using LLMs with chain-of-thought reasoning and few-shot learning. The high quality APIs generated by our tool can further facilitate other relevant research.

Another task related to code APIzation is the generation of API calls~\cite{gu2016deep,huang2018api,zhang2023toolcoder,wei2022clear,wang2023measuring,patil2023gorilla, nguyen2019focus}, which aims to help developers search or generate suitable API calls. 
The goal of this task is to generate a call to an existing API (i.e., the method name and corresponding parameter list) based on a natural language description. 
In recent years, Wei et al.~\cite{wei2022clear} proposed CLEAR, which can recommend reasonable API calls without parameter lists from SO to users based on their questions. 
Wang et at.~\cite{wang2023measuring} and Patil et al.~\cite{patil2023gorilla} used generative models to generate API calls from natural language descriptions respectively. 
Unlike them, our task is to generate reusable APIs, which not only need to generate the method name and parameter list, but also the corresponding method body, making it complicated and challenging task. 

\subsection{Large Language Models for Software Engineering}
LLMs~\cite{zhao2023survey,wang2023software,wang2021codet5,touvron2023llama,wang2022self,chowdhery2022palm,su2023still,xue2023acwrecommender} are increasingly used in software engineering, showing great potential in various tasks. They can be applied through two main methods: fine-tuning and prompt engineering. Fine-tuning adapts LLMs to specific tasks. For example, Codex~\cite{chen2021evaluating} fine-tuned GPT3~\cite{brown2020language} to generate Python functions. 
Thapa et al.~\cite{thapa2022transformer} fine-tuned models like Bert~\cite{devlin2018bert} and GPT2~\cite{radford2019language} to solve software vulnerability detection tasks. 
However, fine-tuning requires high-performance hardware devices, which is expensive and costly for developers with limited computing resources. 
Prompt engineering, on the other hand, uses tailored prompts to leverage LLMs for tasks such as program repair~\cite{paulenhancing,huang2023chain}, code generation~\cite{dong2023self,tan2023copilot}, and test case generation~\cite{siddiq2023exploring,xie2023chatunitest}, without the need for costly hardware. 
Recently, Bareiss et al.~\cite{bareiss2022code} designed specific prompts to solve tasks such as test case generation. 
Li et al.~\cite{li2023codeie} proposed CodeIE and designed corresponding prompts to solve information retrieval tasks. 
To the best of our knowledge, \toolname is the first model to use LLMs with prompt engineering to guide LLMs to solve the APIzation task.

%% file: threat.tex
Several threats to validity are related to our research: 

\noindent\textbf{Threats to internal validity.} 
One threat to internal validity is that the design of prompts can be diverse. For instance, the selection of examples in few-shot learning and the choice of CoTs in chain-of-thought reasoning can both influence model results. While our current prompt may not necessarily be optimal, they have already performed well in the code APIzation task. In the future, we will investigate methods to automatically design high-quality prompts for accomplishing corresponding tasks. 
Another threat is the instability of ChatGPT's output. To reduce the uncertainty of the model's output, we set the corresponding temperature parameter to 0. Nevertheless, the model still has slight instability, which is determined by the implementation of ChatGPT.

\noindent\textbf{Threats to external validity.} 
One threat to external validity is that Code2API is unable to generate APIs for very long Stack Overflow posts, as Chatgpt has a maximum input length of 4096 tokens. 
Considering Chatgpt's support for conversational Q\&A, one potential solution is to employ multiple-round questioning to extract summaries of the question-answer posts to reduce their length. 
These extracted summaries can then be used to replace the original question-answer posts for the APIzation task. 
Another threat is the selection of the evaluation dataset. In this paper, we use the evaluation set from Terragni et al.~\cite{terragni2021apization} to ensure the fairness of the comparison.

\noindent\textbf{Threats to construct validity.} 
One threat to construct validity is the subjectivity and personal bias in the human evaluation process. The preliminary study conducted by the first author has a certain degree of subjectivity. To reduce the subjectivity and personal bias in the manual annotation process, we performed a comprehensive user study in RQ-2 with 18 experienced Java developers. 
The evaluation results are consistent with our preliminary study and the Cohen's kappa coefficient between different evaluation groups further verifies the advantage of our approach. 


%% file: conclusion.tex
This paper aims to automatically solve the APIzation task to facilitate developers in reusing code snippets. To address this task, we employ chain-of-thought reasoning and few-shot in-context learning to guide the LLM to gradually generate a reusable API in a developer-like manner. Extensive evaluations have proven the effectiveness of our approach, and its promising performance and efficiency have led us to develop a practical Google extension to display the corresponding APIs of code snippets when developers browse SO pages, making it easier for them to reuse code snippets. 

\noindent\textbf{Implications.} We propose \toolname for automatically generating reusable APIs for SO code snippets with CoT reasoning and few-shot learning. 
We believe that the implications of \toolname extend far beyond performance gains. 
(1) For software engineering researchers, \toolname proposes a novel way of generating real reusable APIs with LLMs, exploiting the possibilities of using LLMs on a new software engineering task. 
(2) For developers, \toolname provides a Chrome extension tool to help developers use SO code snippets more effectively and efficiently. 
(3) For SO organizers, \toolname has generated two high-quality API datasets and provides a better way to manage these diverse code snippets. Moreover, \toolname can be easily extended to other programming languages. 
In future work, we plan to make high-quality datasets for the APIzation task and fine-tune code LLMs (e.g., CodeLlama~\cite{roziere2023code}) to further enhance the model performance. 
